\documentclass[10pt,aps,prb,amsmath,amssymb,twocolumn,superscriptaddress, notitlepage]{revtex4-2}

\usepackage[bookmarksnumbered,pdfpagelabels=true,plainpages=false,colorlinks=true,linkcolor=blue,citecolor=blue,urlcolor=blue]{hyperref}

\usepackage{graphicx}
\usepackage{bm}
\usepackage{braket}
\usepackage{array}
\usepackage{capt-of}   
\usepackage{ulem} 

\begin{document}
\title{Higher‑order magnetic hopfions in uniaxial chiral magnets}
\author{Vladyslav M. Kuchkin}
\email{vladyslav.kuchkin@uni.lu}
\affiliation{Department of Physics and Materials Science, University of Luxembourg, L-1511 Luxembourg, Luxembourg}
\author{Nikolai~S.~Kiselev}
\affiliation{Peter Gr\"unberg Institute and Institute for Advanced Simulation, Forschungszentrum J\"ulich and JARA, 52425 J\"ulich, Germany}

\makeatother

\begin{abstract}
Chiral magnets host hopfions, but their experimentally observed configurations
are largely limited to hopfion rings, heliknotons,
and their clusters.
Here, we show that uniaxial anisotropy stabilizes a diverse family
of knotted and linked hopfions with zero and arbitrary positive Hopf indices.
We identify their stability regions in the anisotropy--field
parameter space and introduce a systematic construction based
on mapping skyrmion braids into three-dimensional magnetic textures.
We find that the minimum energy of stable hopfions
increases approximately linearly with the Hopf index.
\end{abstract}

\maketitle

\noindent
Magnetic hopfions are three-dimensional (3D) topological solitons characterized by an integer Hopf index $H$~\cite{Whitehead1947}.
They form spatially localized magnetization textures that can
be statically stable and exhibit particle-like properties.
Two-dimensional (2D) magnetic systems support a wide variety of
solitonic textures, including merons~\cite{Komineas2007,Rybakov2025}, skyrmions~\cite{Bogdanov1999Stability}, skyrmion bags~\cite{Rybakov2019,Foster2019},
and chiral kinks~\cite{Kuchkin2020Magnetic,Kuchkin2023Tailed,ku_thesis}.
A similarly rich variety of topological solitons
can be expected in 3D magnetic systems.
The existence and stability of magnetic hopfions have been
investigated particularly in two classes of magnetic systems:
\textit{exchange-frustrated} magnets and \textit{chiral} magnets.
In frustrated magnets, competing exchange interactions can stabilize a broad family of hopfions with distinct knot and link structures and different Hopf indices~\cite{Sutcliffe2017, Rybakov2022}.
Despite these theoretical predictions, experimental evidence for hopfions in frustrated magnets is still lacking.
In contrast, hopfions have already been observed experimentally
in noncentrosymmetric chiral magnets~\cite{Zheng2023,Li2026,Chen2026_2},
where their stability is governed by the interplay of
exchange and Dzyaloshinskii-Moriya interactions (DMI).
Nevertheless, the diversity of spatially localized hopfions
in these experimentally accessible systems remains remarkably limited.

The best-established examples of localized hopfions in chiral magnets are heliknotons -- hopfions embedded in helical background.
Similar textures were originally observed in chiral liquid crystals and subsequently were theoretically predicted to exist in chiral magnets
~\cite{Tai2019,Tai2022,Voinescu2020,Kuchkin2023,Liu2025}.
An isolated heliknoton carries a Hopf index $|H|=1$.
Unlike the skyrmion number of corresponding 2D textures, the Hopf index of these 3D configurations generally depends on the chirality of the spin texture (see Sec.~\ref{sec:hopf-charge} of Ref.~\cite{SupplementalMaterial}).
In right-handed and left-handed chiral magnets with positive and negative DMI constant, we observe stable hopfions with $H=-1$ and $H=1$, respectively~\cite{Chen2026_1}.
Configurations with $|H|>1$ are usually clusters of identical morphology hopfions~\cite{Voinescu2020}.
Metastable hopfions have also been predicted
in a theoretical model of screw chiral magnets
with spatially modulated DMI, although
the experimental realization of such magnetic
systems remains an open question~\cite{Shaju2026}.

Another class of experimentally observed configurations consists of hopfion rings linked to skyrmion strings~\cite{ Chen2026_1, Zheng2023}.
Unlike isolated hopfions, these structures are localized
in only two spatial directions and are characterized
by three topological indices.
Consequently, despite the variety of three-dimensional
magnetic textures observed in chiral magnets
~\cite{Tang2021,Zheng2021,Zheng2023,Zhang2024},
the existence of a broad family of isolated knotted
and linked hopfions has remained unresolved.

This restriction raises two closely related questions.
Can chiral magnets stabilize isolated hopfions with
substantially different knot and link structures,
and which material parameters determine their stability?
Equally important is how to construct such configurations,
since identifying suitable initial magnetization textures
is a major obstacle to their numerical investigation.

Here, we show that uniaxial magnetic anisotropy stabilizes
a broad family of spatially localized hopfions with
diverse knot and link structures in chiral magnets.
To construct these configurations, we employ braid closure,
mapping periodically braided skyrmion strings onto
3D magnetization textures.

\begin{figure*}[tb!]
\centering
\includegraphics[width=18cm]{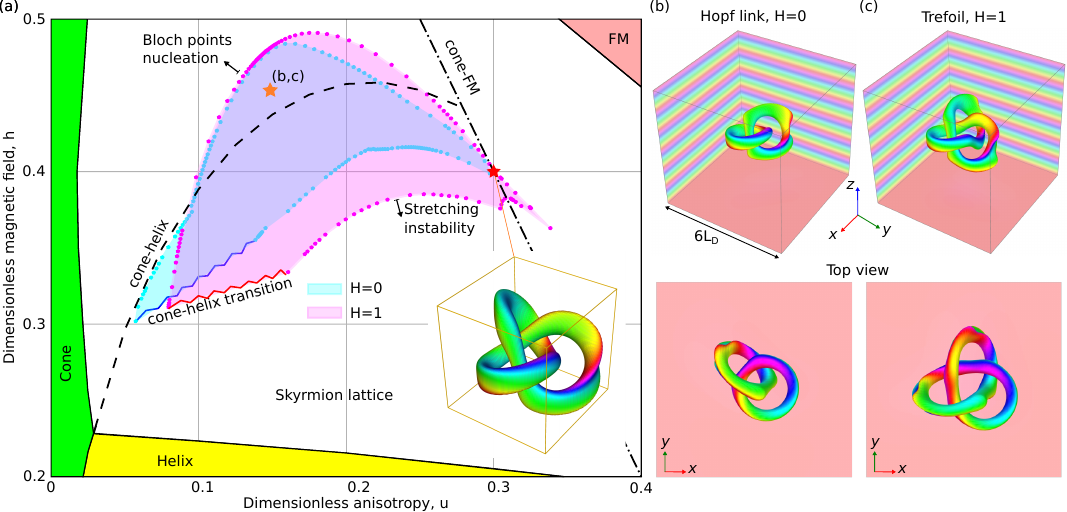}
\caption{~\small \textbf{Hopfion stability diagram}. 
(a): Phase diagram of a cubic chiral magnet with anisotropy, reproduced from Ref.~\cite{Wilson2014}; the ferromagnet (FM), skyrmion-lattice, cone, and helix phases are highlighted.
Cyan and magenta regions indicate the stability ranges of
the Hopf link and refoil hopfion, respectively,
as determined from the numerical data points. 
The inset shows a stable trefoil hopfion in the FM background,
with a box indicating its orientation within a larger simulation domain.
(b) and (c): 3D spin textures (upper row) and top view (bottom row) for the Hopf link and the Trefoil hopfion at $(u, h)=(0.25, 0.45)$, given by the isosurface $m_{z}=0$ and color-coded by $m_{x}$ and $m_{y}$~\cite{Savchenko2022}.
} 
\label{Fig1}
\end{figure*}

\textit{Micromagnetic model} -- A standard model for chiral magnets with isotropic exchange, $w_\mathrm{ex}=\left(\nabla \mathbf{m}\right)^{2}$, and bulk type DMI, $w_\mathrm{DM}=\mathbf{m}\cdot\nabla\times\mathbf{m}$, can be written as follows:
\begin{equation}
    \mathcal{E}_{0}=\int \mathrm{d}V\left[\mathcal{A}w_\mathrm{ex} + \mathcal{D}w_\mathrm{dmi}+w_{\rm pot}\right],\label{Energy_micro}
\end{equation}
where $\mathcal{A}$ and $\mathcal{D}$ are strengths of exchange and DMI, respectively. 
The potential energy density is
$w_{\rm pot}=-\frac{1}{2}M_\mathrm{s}\mathbf{B}_\mathrm{d}
\cdot\mathbf{m}+M_\mathrm{s}B(1-m_{z})
+\mathcal{K}(1-m_{z}^{2})$,
where $\mathbf{B}_\mathrm{d}$ is the demagnetizing field
and $B$ is the externally applied magnetic field
directed along the $z$-axis.
The constant $\mathcal{K}$ describes uniaxial magnetocrystalline anisotropy, with the $z$-axis as the easy axis ($\mathcal{K}>0$) or hard axis ($\mathcal{K}<0$).
We consider bulk chiral magnets with periodic boundary
conditions in all three spatial directions and account for
magnetostatic interactions through additional effective easy-plane anisotropy, $\mathcal{K}_{d} = -\mu_{0}M_{s}/2$ ~\cite{Zheng2021}.

Following the standard approach~\cite{Bogdanov_89,Bogdanov_1994},
we introduce the characteristic length
$L_{\rm D}=4\pi\mathcal{A}/\mathcal{D}$,
corresponding to the zero-field spin-spiral period,
and the characteristic magnetic field
$B_{\rm D}=\mathcal{D}^{2}/(2\mathcal{A}M_{\rm s})$.
These scales define the dimensionless coordinates
$\mathbf{r}\mapsto\mathbf{r}/L_{\rm D}$,
energy $\mathcal{E}=\mathcal{E}_{0}/(2\mathcal{A}L_{\rm D})$,
magnetic field $h=B/B_{\rm D}$,
and anisotropy $u=K_z/(M_{\rm s}B_{\rm D})-\mu_{0}/2B_{D}$.
In these units, the potential energy density takes the form
$w=4\pi^{2}[h(1-m_z)+u(1-m_z^2)]$.
For numerical calculations, we discretize a cubic
simulation box of side length $12L_{\rm D}$
into $256^3$ cells.
The energy functional is evaluated using
higher-order finite-difference approximations
(see Sec.~\ref{sec:fin-diff} of Ref.~\cite{SupplementalMaterial}).
We obtain equilibrium magnetization configurations
by minimizing the discretized energy functional
using the conjugate-gradient method~\cite{Rybakov2019}.
The Hopf index $H$ is calculated using the method
introduced in Ref.~\cite{Chen2026_2},
with implementation details provided in
Sec.~\ref{sec:hopf-charge} of Ref.~\cite{SupplementalMaterial}.

\textit{Hopfion stability diagram} --
To determine the conditions under which higher-order hopfions
can exist, we first examine two representative configurations:
the Hopf link ($H=0$) and the trefoil knot ($H=1$).
The corresponding stability regions are shown in
Fig.~\ref{Fig1}(a), together with the equilibrium phase diagram
of a chiral magnet with uniaxial anisotropy,
reproduced from Ref.~\cite{Wilson2014}.
The diagram contains four equilibrium phases:
the ferromagnetic (FM) state with $\mathbf{m}=\mathbf{e}_{z}$,
the skyrmion lattice, and two spiral states:
the cone with its propagation vector $\mathbf{q}$
parallel to the field and the helix with $\mathbf{q}$
perpendicular to the field.
Both hopfions are stable local energy minima in the
metastable conical background, where the skyrmion lattice
is the ground state.
Their relaxed spin textures at $(u,h)=(0.25,0.45)$
are shown in Figs.~\ref{Fig1}(b) and \ref{Fig1}(c).
At higher fields, the hopfions become unstable through
Bloch-point nucleation~\cite{Doring1968},
whereas at lower fields they undergo various
stretching deformations.
At sufficiently low fields and anisotropies,
the conical background itself becomes unstable,
and the hopfion can initiate its transformation
into the helical state.
The diagram also includes the cone-helix and cone-FM equal-energy lines.
Unlike the Hopf link, which requires a conical background, the trefoil hopfion can be stable even in the FM background within a narrow parameter range.
The uniform FM background is particularly favorable for studying the intrinsic dynamical modes of hopfions.
Unlike the conical background, it is not excited by magnetic field pulses applied parallel to its magnetization, allowing the soliton dynamics to be probed without simultaneously exciting the background.
A similar approach has been used to investigate the dynamics of 2D chiral skyrmions\cite{Kuchkin2025_v3}.

\begin{figure*}[tb!]
\centering
\includegraphics[width=18cm]{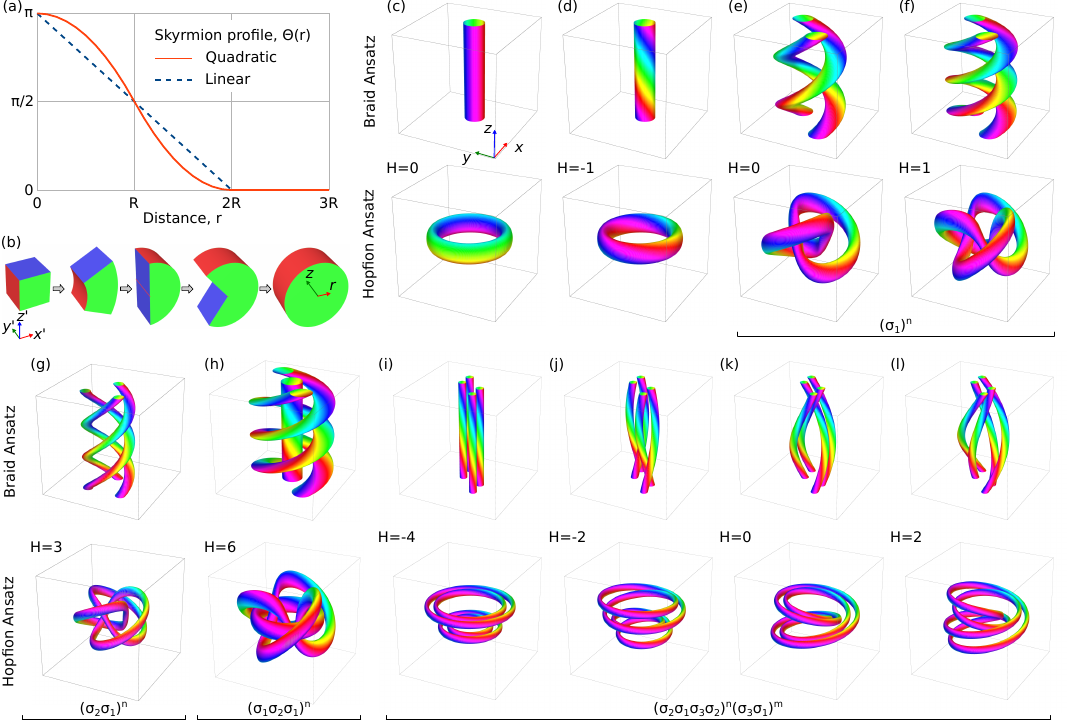}
\caption{~\small \textbf{Hopfion ansatz construction}.
(a) Skyrmion profiles defined by quadratic
[Eq.~\eqref{sk_ansatz}] and linear functions of $r$.
(b) Illustration of the transformation
defined by Eq.~\eqref{Tranformation}.
(c) and (d) Examples of applying the transformation
to ordinary and twisted skyrmion strings,
respectively, yielding toroidal hopfions.
(e) and (f) Construction of a Hopf link ($n=2$)
and a trefoil knot ($n=3$) from two braided
skyrmion strings with braid word $\sigma_1^n$.
(g) and (h) Examples of three-strand skyrmion
braids with words $(\sigma_2\sigma_1)^3$
and $(\sigma_1\sigma_2\sigma_1)^3$, respectively.
(i)--(l) Application of the transformation
to four-strand skyrmion braids with braid word
$(\sigma_2\sigma_1\sigma_3\sigma_2)^n
(\sigma_3\sigma_1)^m$.
Panel (i) shows the unbraided case ($n=m=0$).
Panels (j) and (k) show pairwise braids
with $(n,m)=(0,1)$ and $(1,0)$, respectively.
Panel (l) shows the simplest case
with $(n,m)=(1,1)$ for this braid word.
}
\label{Fig2}
\end{figure*}

\textit{Braid-to-hopfion mapping} --
To construct hopfions with different knot and link
structures, we map periodic skyrmion braids
onto spatially localized 3D magnetization textures.
In bulk samples, magnetic skyrmions can form strings, which may also be braided~\cite{Zheng2021}.
Skyrmion braiding has also been shown
to produce nontrivial spacetime topology
associated with magnetic hopfions~\cite{Knapman2024}.

We describe these structures using piecewise-defined
ansätze for individual skyrmion strings
[see Fig.~\ref{Fig2}(a)].
The initial magnetization field
$\mathbf{m}^{\prime}(\mathbf{r}^{\prime})$
is defined in a cubic domain $[0,L]^3$
with periodic boundary conditions along the $z$-axis,
while the background magnetization is uniform,
$\mathbf{m}^{\prime}=(0,0,1)$.

Then we perform the following transformation [Fig.~\ref{Fig2}(b)] for new magnetization field $\mathbf{m}(x,y,z)$ in a system of size $2L\times 2L \times L$, which for convenience is given in cylindrical coordinates:
\begin{eqnarray}
    \!\!r = x^{\prime},  \phi = \dfrac{2\pi}{L} z^{\prime}, z=y^{\prime}, 
    \quad \mathbf{m}(\mathbf{r}) = \mathbf{m}^{\prime}(\mathbf{r}^{\prime})\,,\label{Tranformation}
\end{eqnarray}
where for $r>L$, we set $\mathbf{m}=(0,0,1)$.
As illustrated in Fig.~\ref{Fig2}(b), this transformation maps each $z^{\prime}$-plane onto a plane of constant $\phi$,
while preserving the local magnetization direction.
In the case of a single skyrmion string, \eqref{Tranformation} allows us to get a toroidal hopfion texture with $H=0$, as demonstrated in Fig.~\ref{Fig2}(c). 
To get a non-trivial toroidal hopfion, we need to apply \eqref{Tranformation} to a twisted skyrmion string~\cite{Saji2025} carrying an additional $z$-dependent phase [see Fig.~\ref{Fig2}(d)].
This case corresponds to a heliknoton $H=-1$ and was already studied previously~\cite{Voinescu2020, Kuchkin2023}.

Hopfions with more complex knot and link structures
can be constructed by applying the transformation~\eqref{Tranformation} to skyrmion braids [Fig.~\ref{Fig2}(e)--(l)].
In chiral magnets, the initial skyrmion strings
must have the same type of twist~\cite{Leonov2016}
to obtain suitable initial configurations for hopfions.
We therefore construct the corresponding ansatz as follows. 
First we define $I$ braiding functions, $x^{\prime}=f_{i}(z^{\prime}), y^{\prime} = g_{i}(z^{\prime}), i=\overline{1,I}$ describing skyrmion core positions, domains of which do not overlap, i.e. if $A_{i}=\{(x^{\prime}, y^{\prime}, z^{\prime}) : \sqrt{\left(x^{\prime}\!-\!f_{i}\right)^{2} \!+\! \left(y^{\prime}\!-\!g_{i}\right)^{2}} < 2R\}$, $\forall i\neq j$, $A_{i}\cap A_{j}= \varnothing$, and $R$ has a sense of skyrmion radius.
Then parameterizing magnetization $\mathbf{m}^{\prime}$ with spherical angles $\Theta^{\prime}, \Phi^{\prime}$, as $\mathbf{m}^{\prime}=\left(\sin\Theta^{\prime}\cos\Phi^{\prime}, \sin\Theta^{\prime}\sin\Phi^{\prime}, \cos\Theta^{\prime}\right)$, we can write the ansatz for the angles [see Fig.~\ref{Fig2}(a)] in each set $A_{i}$, i.e. $r_{i}^{\prime}<2R$:
\begin{align}
   \!\!\!\Theta^{\prime}_{i}=\begin{cases}
       \pi\left(1-\dfrac{r^{\prime 2}_{i}}{2R^{2}}\right), r^{\prime}_{i}<R, \\
       2\pi\left(1-\dfrac{r^{\prime}_{i}}{2R}\right)^{2}\!\!, r^{\prime}_{i}\geq R,
   \end{cases}\!\!\!\!\!\! \Phi^{\prime}_{i} = \phi^{\prime}_{i} + \dfrac{\pi}{2} +  \dfrac{2\pi z^{\prime}}{L},\label{sk_ansatz}  
\end{align}
where the local cylindrical coordinates are
$r_i'=\sqrt{(x'\!-\!f_i)^2\!+\!(y'\!-\!g_i)^2}$
and
$\phi_i'\!=\!\operatorname{atan2}(y'\!-\!g_i,x'\!-\!f_i)$.
Outside the domains $A_i$, the magnetization is uniform, $\mathbf{m}'=(0,0,1)$.
For micromagnetic simulations on a discrete Cartesian grid, the transformed magnetization field is obtained by interpolation from a cylindrical grid, as described in Sec.~\ref{sec:Interpolation} of Ref.~\cite{SupplementalMaterial}.

The final step in constructing the hopfion ansatz is to introduce the conical background magnetization $\mathbf{m}_{c}$ with angles $\Phi_{c}(z)$ and $\Theta_{c}=\mathrm{const}$.
This is achieved by applying the following rotations to the magnetization field $\mathbf{m}(\mathbf{r})$:
\begin{equation}
    \mathcal{R}_{z}\left(\Phi_{c}\right)\mathcal{R}_{x}\left(\Theta_{c}\right)\mathcal{R}_{z}^{-1}\left(\Phi_{c}\right)\mathbf{m}(\mathbf{r}),\label{uniform_to_cone}
\end{equation}
where $\mathcal{R}_{a}\left(\alpha\right)$ is $3\times 3$ rotational matrix of angle $\alpha$ about $a$-axis.
For large cone angles $\Theta_c$, this transformation
may introduce distortions near the hopfion cores
($m_z=-1$).
To reduce this effect, we replace $\Theta_c$ with
$\Theta_c^{*}=\Theta_c(m_z+1)^2/4$.

\textit{Morphological classification of hopfions} -- The Hopf index $H$ alone is insufficient to distinguish the variety of hopfion configurations.
As in the case of 2D magnetic skyrmions, different configurations may have the same topological index while being separated by finite energy barriers.
Such hopfion configurations can differ in their spatial structure, size, and energy.
To distinguish these configurations, we introduce an additional classification based on braid words~\cite{Adams2004KnotBook, Murasugi1996KnotTheory}, which describe the braiding of skyrmion strings used to construct the corresponding hopfions.

A braid word describes the sequence of exchanges between neighboring skyrmion strings.
The symbol $\sigma_i$ denotes an exchange of the $(i+1)$-th and $i$-th strings, while $\sigma_i^{-1}$ denotes the exchange in the opposite direction.
The order of these exchanges determines the braiding pattern, and a power $n$ indicates that the corresponding sequence is repeated $n$ times.
The explicit braiding functions used to construct
the initial magnetization fields are given in
Sec.~\ref{sec:Interpolation}
of Ref.~\cite{SupplementalMaterial}.

\begin{figure*}[tb!]
\centering
\includegraphics[width=18cm]{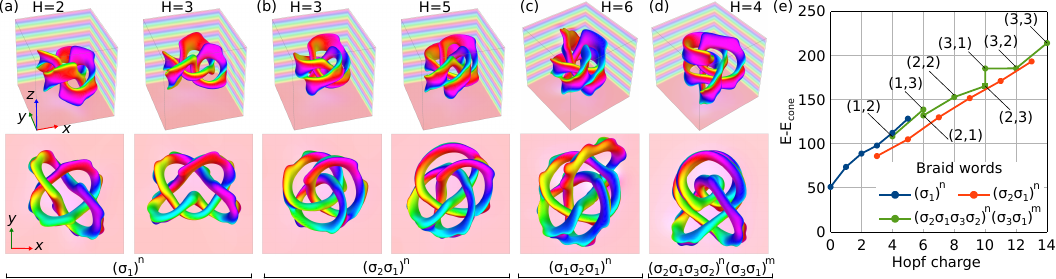}
\caption{~\small \textbf{Examples of higher-order hopfions}.
(a)--(d) Hopfion configurations shown in a cropped
box of size $(6l)^3$ (upper row) and their
corresponding top views (bottom row).
The configurations in (a) and (b) are stabilized
in a simulation domain of volume $V=(12l)^3$,
while those in (c) and (d) are stabilized
in a domain of volume $V=(24l)^3$,
at $(u,h)=(0.2,0.43)$.
(e) Energy of different hopfion configurations
as a function of their Hopf index, relative
to the energy of the conical phase,
$E_{\mathrm{cone}}=4\pi^2V
\left[2(h+u)-1-{h^2}/(1-2u)\right]$.
The values of $(n,m)$ corresponding to the
braid word
$(\sigma_2\sigma_1\sigma_3\sigma_2)^n
(\sigma_3\sigma_1)^m$
are indicated.
}
\label{Fig3}
\end{figure*}

For two skyrmion strings, the braid word
takes the form $\sigma_1^n$.
We find stable hopfions for $n\geq 2$,
with Hopf index $H=n-2$.
The cases $n=2$ and $n=3$ correspond to
the Hopf link and the trefoil knot,
respectively [Figs.~\ref{Fig1}(b) and \ref{Fig1}(c)].
Examples with $n=4$ and $n=5$ are shown
in Fig.~\ref{Fig3}(a).

For three or more skyrmion strings, the braid group
is non-Abelian, allowing for more complex
braid structures.
Here, we consider two families of braid words:
$(\sigma_2\sigma_1)^n$ and
$(\sigma_1\sigma_2\sigma_1)^n$
[Figs.~\ref{Fig2}(f) and \ref{Fig2}(g)].
These families represent different braiding
patterns and produce hopfions with
$H=2n-3$ and $H=3n-3$, respectively.
In both cases, stable configurations
are found for $n\geq 3$
[Figs.~\ref{Fig3}(b) and \ref{Fig3}(c)].

We also construct hopfions from four
skyrmion strings using the braid word
$(\sigma_2\sigma_1\sigma_3\sigma_2)^n
(\sigma_3\sigma_1)^m$,
which gives $H=4n+2m-4$
[Figs.~\ref{Fig2}(i)--(l)].
Stable configurations are found for
$m+n>2$.
An example with $n=1$ and $m=2$
is shown in Fig.~\ref{Fig3}(d).
These examples demonstrate the diversity
of hopfion structures that can be obtained
from skyrmion braids.
Additional configurations are presented
in Sec.~\ref{sec:hopfion_diversity}
of Ref.~\cite{SupplementalMaterial}.

\textit{Energy vs. topological charge} --
The variety of stable hopfions allows us to examine how their energy depends on the Hopf index $H$.
As shown in Fig.~\ref{Fig3}(e), the lowest energy among the configurations found for each $H$ increases approximately linearly with $H$, despite substantial energy differences between hopfions with the same topological charge.
This behavior differs from the $\mathcal{E}\propto |H|^{3/4}$ scaling reported for hopfions in frustrated magnets~\cite{Rybakov2022}.
Another distinctive feature is the existence of stable hopfions with $H=0$.
Since these states have the same Hopf index as the background, their formation is not topologically forbidden and may occur without appearance of magnetic singularities.
This suggests a possible mechanism for their thermal nucleation, similar to processes discussed for 2D chiral magnets~\cite{Kuchkin2022, Kuchkin2022_v2}.
Nucleating hopfions with $H>0$ may require stronger excitations, such as current or laser pulses, as recently demonstrated for heliknotons~\cite{Chen2026_1, Chen2026_2, Li2026}.

An important consequence of the morphological
diversity is that hopfions with the same
Hopf index can represent distinct
metastable configurations.
For example, the configurations with
$H=3$ in Figs.~\ref{Fig3}(a) and
\ref{Fig3}(b), constructed from
$\sigma_1^5$ and $(\sigma_2\sigma_1)^3$,
have different structures despite
their identical Hopf indices.
Although these configurations are
topologically equivalent, they may be
separated by an energy barrier.
Identifying possible transitions between
such states remains an open question.

\textit{Conclusions} --
We have demonstrated the existence of a wide variety
of metastable hopfions with different knot and link
structures in bulk chiral magnets with uniaxial magnetic anisotropy.
We determined the stability regions of two
representative configurations, which provide
guidance for identifying suitable parameters
for other hopfions.
The approach we present is not limited to magnetism only, and we believe it will be helpful in constructing and searching for hopfion textures in various physical systems: ferroelectrics~\cite{Lukyanchuk2020, GmezOrtiz2024}, Bose-Einstein condensates~\cite{Bisset2015, Zhao2025}, superfluid $^{3}$He~\cite{Volovik1977ParticleLike}, optical~\cite{Lin2025, Yifan2026}, etc.

An important question is how the uniaxial
anisotropy required to stabilize these hopfions
can be realized experimentally.
One possibility is magnetoelastic coupling,
through which uniaxial strain can induce
additional magnetic anisotropy.
This suggests that mechanical strain could
be used to control the stability of
knotted and linked magnetic textures.
Another possibility is to use shape anisotropy
in elongated samples.
Such samples can be fabricated from B20-type
chiral magnets or Co--Zn--Mn alloys in the
form of wires or needles using focused ion beam milling.

\textit{Acknowledgments}
VMK acknowledges the financial support from the European Union’s Horizon Europe research and innovation programme under the Marie Sk{\l}odowska-Curie grant agreement No.~101203692 (QUANTHOPF), and the Luxembourg National Research Fund under Grants C22/MS/17415246/DeQuSky and AFR/23/17951349.
NSK acknowledges funding from the European Research Council under the European Union's Horizon 2020 Research and Innovation Programme (Grant No.~856538 - project ``3D MAGiC'') and the Deutsche Forschungsgemeinschaft (Project-ID 405553726 -- TRR~270; Priority Programme SPP 2137; Project No.~403502830).

\textit{Data Availability} -- The Mumax3 scripts and hopfions initial states are published on Zenodo~\cite{mumax3_scripts}.

\bibliography{main_Arxiv}

@article{BERG1981412,
author = {B. Berg and M. Lüscher},
title = {Definition and statistical distributions of a topological number in the lattice {$O(3)$} $\sigma$-model},
journal = {Nucl. Phys. B},
volume = {190},
number = {2},
pages = {412-424},
year = {1981},
issn = {0550-3213},
doi = {10.1016/0550-3213(81)90568-X},
url = {https://www.sciencedirect.com/science/article/pii/055032138190568X}
}

@article{Shaju2026,
  title = {Hopfions in screw chiral magnets},
  volume = {14},
  ISSN = {2166-532X},
  url = {http://dx.doi.org/10.1063/5.0327533},
  DOI = {10.1063/5.0327533},
  number = {5},
  journal = {APL Materials},
  publisher = {AIP Publishing},
  author = {Shaju,  Sandra Chulliparambil and Azhar,  Maria and Everschor-Sitte,  Karin},
  pages = {051111},
  year = {2026},
  month = May 
}

@misc{SupplementalMaterial,
    note = {See Supplemental Material at [URL inserted by publisher] contains details of the numerical methods, interpolation
procedures, Hopf index calculation, and an extended set of Hopfions.
    }
}

@article{Lukyanchuk2020,
  title = {Hopfions emerge in ferroelectrics},
  volume = {11},
  ISSN = {2041-1723},
  url = {http://dx.doi.org/10.1038/s41467-020-16258-w},
  number = {1},
  journal = {Nat. Commun.},
  publisher = {Springer Science and Business Media LLC},
  author = {Luk’yanchuk,  I. and Tikhonov,  Y. and Razumnaya,  A. and Vinokur,  V. M.},
  year = {2020},
  month = May 
}

@article{Bisset2015,
  title = {Robust vortex lines, vortex rings, and hopfions in three-dimensional Bose-Einstein condensates},
  author = {Bisset, R. N. and Wang, Wenlong and Ticknor, C. and Carretero-Gonz{\'a}lez, R. and Frantzeskakis, D. J. and Collins, L. A. and Kevrekidis, P. G.},
  journal = {Phys. Rev. A},
  volume = {92},
  issue = {6},
  pages = {063611},
  numpages = {7},
  year = {2015},
  month = {Dec},
  publisher = {American Physical Society},
  doi = {10.1103/PhysRevA.92.063611},
  url = {https://link.aps.org/doi/10.1103/PhysRevA.92.063611}
}

@article{Volovik1977ParticleLike,
  title={Particle-like solitons in superfluid $^3${H}e phases},
  author={Volovik, G E and Mineev, V P},
  journal={Zh. Eksp. Teor. Fiz},
  volume={73},
  pages={767--773},
  year={1977},
  url={https://jetp.ras.ru/cgi-bin/dn/e_046_02_0401.pdf}
}

@article{Lin2025,
  title = {Space-Time Optical Hopfion Crystals},
  author = {Lin, Wenbo and Mata-Cervera, Nilo and Ota, Yasutomo and Shen, Yijie and Iwamoto, Satoshi},
  journal = {Phys. Rev. Lett.},
  volume = {135},
  issue = {8},
  pages = {083801},
  numpages = {6},
  year = {2025},
  month = {Aug},
  publisher = {American Physical Society},
  doi = {10.1103/hh5s-cprt},
  url = {https://link.aps.org/doi/10.1103/hh5s-cprt}
}

@misc{Yifan2026,
  doi = {10.48550/ARXIV.2611.13476},
  url = {https://arxiv.org/abs/2611.13476},
  author = {Zhang,  Yifan and Dev,  Vasu and Shen,  Yijie},
  title = {Hopfion Superlattices in 3D Twisted Moiré Light Fields},
  publisher = {arXiv},
  year = {2026},
  copyright = {arXiv.org perpetual,  non-exclusive license}
}

@misc{Zhao2025,
  doi = {10.48550/ARXIV.2507.10910},
  url = {https://arxiv.org/abs/2507.10910},
  author = {Zhao,  Zibin and Li,  Guilong and Luo,  Huanbo and Liu,  Bin and Chen,  Guihua and Malomed,  Boris A. and Li,  Yongyao},
  title = {Stable hopfions in trapped quantum droplets},
  publisher = {arXiv},
  year = {2025},
  copyright = {Creative Commons Attribution 4.0 International}
}

@article{GmezOrtiz2024,
  title = {Liquid-Crystal-Like Dynamic Transition in Ferroelectric-Dielectric Superlattices},
  volume = {133},
  ISSN = {1079-7114},
  url = {http://dx.doi.org/10.1103/PhysRevLett.133.066801},
  number = {6},
  journal = {Phys. Rev. Lett.},
  publisher = {American Physical Society (APS)},
  author = {Gómez-Ortiz,  Fernando and Graf,  Mónica and Junquera,  Javier and Íñiguez-González,  Jorge and Aramberri,  Hugo},
  year = {2024},
  month = Aug 
}

@article{Azhar2026,
  title = {3D magnetic textures with mixed topology: Unlocking the tunable Hopf index},
  volume = {21},
  ISSN = {2542-4653},
  url = {http://dx.doi.org/10.21468/SciPostPhys.21.1.023},
  DOI = {10.21468/scipostphys.21.1.023},
  number = {1},
  journal = {SciPost Physics},
  publisher = {Stichting SciPost},
  author = {Azhar,  Maria and Shaju,  Sandra C. and Knapman,  Ross and Pignedoli,  Alessandro and Everschor-Sitte,  Karin},
  pages = {023},
  year = {2026},
  month = July 
}

@article{Whitehead1947,
  title = {An Expression of Hopf’s Invariant as an Integral},
  volume = {33},
  ISSN = {1091-6490},
  url = {http://dx.doi.org/10.1073/pnas.33.5.117},
  DOI = {10.1073/pnas.33.5.117},
  number = {5},
  journal = {Proc. Natl. Acad. Sci. U.S.A.},
  publisher = {National Academy of Sciences},
  author = {Whitehead,  J. H. C.},
  year = {1947},
  month = May,
  pages = {117–123}
}

@article{Knapman2024,
  title = {Spacetime magnetic hopfions from internal excitations and braiding of skyrmions},
  volume = {7},
  ISSN = {2399-3650},
  url = {http://dx.doi.org/10.1038/s42005-024-01628-3},
  DOI = {10.1038/s42005-024-01628-3},
  number = {1},
  pages = {151},
  journal = {Communications Physics},
  publisher = {Springer Science and Business Media LLC},
  author = {Knapman,  Ross and Tausendpfund,  Timon and Díaz,  Sebastián A. and Everschor-Sitte,  Karin},
  year = {2024},
  month = May 
}

@article{Knapman2025,
  title = {Numerical calculation of the Hopf index for three-dimensional magnetic textures},
  author = {Knapman, R. and Azhar, M. and Pignedoli, A. and Gallard, L. and Hertel, R. and Leliaert, J. and Everschor-Sitte, K.},
  journal = {Phys. Rev. B},
  volume = {111},
  issue = {13},
  pages = {134408},
  numpages = {7},
  year = {2025},
  month = {Apr},
  publisher = {American Physical Society},
  doi = {10.1103/PhysRevB.111.134408},
  url = {https://link.aps.org/doi/10.1103/PhysRevB.111.134408}
}

@article{Li2026,
  title = {Electrically writing a magnetic heliknoton in a chiral magnet},
  volume = {25},
  ISSN = {1476-4660},
  url = {http://dx.doi.org/10.1038/s41563-025-02450-0},
  DOI = {10.1038/s41563-025-02450-0},
  number = {6},
  journal = {Nat. Mater.},
  publisher = {Springer Science and Business Media LLC},
  author = {Li,  Long and Song,  Dongsheng and Wang,  Weiwei and Kong,  Lingyao and Zhang,  Shuisen and Wang,  Ning and Zhang,  Shilei and Tian,  Mingliang and Zang,  Jiadong and Liu,  Yizhou and Du,  Haifeng},
  year = {2026},
  month = Jan,
  pages = {950–955}
}

@article{Sutcliffe2017,
  title = {Skyrmion Knots in Frustrated Magnets},
  author = {Sutcliffe, Paul},
  journal = {Phys. Rev. Lett.},
  volume = {118},
  issue = {24},
  pages = {247203},
  numpages = {4},
  year = {2017},
  month = {Jun},
  publisher = {American Physical Society},
  doi = {10.1103/PhysRevLett.118.247203},
  url = {https://link.aps.org/doi/10.1103/PhysRevLett.118.247203}
}

@misc{Adams2004KnotBook,
  author       = {Adams, Colin C.},
  title        = {The Knot Book: An Elementary Introduction to the Mathematical Theory of Knots},
  howpublished = {American Mathematical Society, Providence, RI},
  year         = {2004},
}

@misc{Murasugi1996KnotTheory,
  author       = {Murasugi, Kunio},
  title        = {Knot Theory and Its Applications},
  howpublished = {Birkh{\"a}user, Boston},
  year         = {1996},
}

@article{Bogdanov_89,
  author  = {A. N. Bogdanov and D. A. Yablonskii},
  journal = {Sov. Phys. JETP},
  number  = {101},
  title   = {Thermodynamically stable ``vortices'' in magnetically ordered crystals. The mixed state of magnets},
  url     = {http://www.jetp.ac.ru/cgi-bin/e/index/e/68/1/p101?a=list},
  volume  = {68},
  year    = {1989},
  pages     = {101}
}

@article{Bogdanov_1994,
  author    = {A. Bogdanov and A. Hubert},
  doi       = {10.1002/pssb.2221860223},
  journal   = {Phys. Status Solidi B},
  month     = dec,
  number    = {2},
  pages     = {527--543},
  publisher = {Wiley},
  title     = {The Properties of Isolated Magnetic Vortices},
  url       = {https://doi.org/10.1002/pssb.2221860223},
  volume    = {186},
  year      = {1994}
}

@article{Wilson2014,
  title = {Chiral skyrmions in cubic helimagnet films: The role of uniaxial anisotropy},
  volume = {89},
  ISSN = {1550-235X},
  url = {http://dx.doi.org/10.1103/PhysRevB.89.094411},
  DOI = {10.1103/physrevb.89.094411},
  number = {9},
  journal = {Phys. Rev. B},
  publisher = {American Physical Society (APS)},
  author = {Wilson,  M. N. and Butenko,  A. B. and Bogdanov,  A. N. and Monchesky,  T. L.},
  year = {2014},
  month = Mar,
  pages     = {094411}
}

@article{Tai2022,
  title = {Geometric transformation and three-dimensional hopping of {Hopf} solitons},
  volume = {13},
  ISSN = {2041-1723},
  url = {http://dx.doi.org/10.1038/s41467-022-30494-2},
  pages = {2986},
  journal = {Nat. Commun.},
  publisher = {Springer Science and Business Media LLC},
  author = {Tai,  Jung-Shen B. and Wu,  Jin-Sheng and Smalyukh,  Ivan I.},
  year = {2022},
  month = May 
}

@article{Fornberg1988,
  title = {Generation of finite difference formulas on arbitrarily spaced grids},
  volume = {51},
  ISSN = {0025-5718},
  url = {http://dx.doi.org/10.1090/S0025-5718-1988-0935077-0},
  number = {184},
  journal = {Math. Comput.},
  publisher = {American Mathematical Society (AMS)},
  author = {Fornberg,  Bengt},
  year = {1988},
  pages = {699–706}
}

@article{Savchenko2022,
  title = {Chiral standing spin waves in skyrmion lattice},
  volume = {10},
  ISSN = {2166-532X},
  url = {http://dx.doi.org/10.1063/5.0097651},
  DOI = {10.1063/5.0097651},
  number = {7},
  journal = {APL Mater.},
  publisher = {AIP Publishing},
  author = {Savchenko,  Andrii S. and Kuchkin,  Vladyslav M. and Rybakov,  Filipp N. and Bl\"{u}gel,  Stefan and Kiselev,  Nikolai S.},
  year = {2022},
  month = {July},
  pages     = {071111}
}

@article{Zheng2023,
  title = {Hopfion rings in a cubic chiral magnet},
  volume = {623},
  ISSN = {1476-4687},
  url = {http://dx.doi.org/10.1038/s41586-023-06658-5},
  DOI = {10.1038/s41586-023-06658-5},
  number = {7988},
  journal = {Nature},
  publisher = {Springer Science and Business Media LLC},
  author = {Zheng,  Fengshan and Kiselev,  Nikolai S. and Rybakov,  Filipp N. and Yang,  Luyan and Shi,  Wen and Bl\"{u}gel,  Stefan and Dunin-Borkowski,  Rafal E.},
  year = {2023},
  month = Nov,
  pages = {718–723}
}

@article{Kuchkin2023Tailed,
  title = {Tailed skyrmions—An obscure branch of magnetic solitons},
  author = {Kuchkin, Vladyslav M. and Kiselev, Nikolai S. and Rybakov, Filipp N. and Bessarab, Pavel F.},
  journal = {Front. Phys.},
  volume = {11},
  pages = {1171079},
  year = {2023},
  month = {apr},
  publisher = {Frontiers Media SA},
  doi = {10.3389/fphy.2023.1171079},
  url = {https://www.frontiersin.org/articles/10.3389/fphy.2023.1171079}
}

@article{Rybakov2025,
  title = {Topological invariants of vortices,  merons,  skyrmions,  and their combinations in continuous and discrete systems},
  volume = {111},
  ISSN = {2469-9969},
  url = {http://dx.doi.org/10.1103/PhysRevB.111.134417},
  number = {13},
  journal = {Phys. Rev. B},
  publisher = {American Physical Society (APS)},
  author = {Rybakov,  Filipp N. and Eriksson,  Olle and Kiselev,  Nikolai S.},
  year = {2025},
  month = Apr,
  pages     = {134417}
}

@article{Komineas2007,
  title = {Rotating Vortex Dipoles in Ferromagnets},
  volume = {99},
  ISSN = {1079-7114},
  url = {http://dx.doi.org/10.1103/PhysRevLett.99.117202},
  number = {11},
  journal = {Phys. Rev. Lett.},
  publisher = {American Physical Society (APS)},
  author = {Komineas,  S.},
  year = {2007},
  month = {Sept},
  pages     = {117202}
}

@book{ku_thesis,
  url = {https://publications.rwth-aachen.de/record/855848},
  author = {Kuchkin,  Vladyslav},
  title = {Diversity of chiral magnetic solitons},
  journal = {Dissertation},
  volume = {RWTH Aachen University},
  pages = {2022},
  publisher = {RWTH Aachen University},
  year = {2022}
}

@article{Kuchkin2020Magnetic,
  title = {Magnetic skyrmions, chiral kinks, and holomorphic functions},
  author = {Kuchkin, Vladyslav M. and Barton-Singer, Bruno and Rybakov, Filipp N. and Bl{\"u}gel, Stefan and Schroers, Bernd J. and Kiselev, Nikolai S.},
  journal = {Phys. Rev. B},
  volume = {102},
  issue = {14},
  pages = {144422},
  numpages = {17},
  year = {2020},
  month = {Oct},
  publisher = {American Physical Society},
  doi = {10.1103/PhysRevB.102.144422},
  url = {https://link.aps.org/doi/10.1103/PhysRevB.102.144422}
}

@article{Bogdanov1999Stability,
  title = {The stability of vortex-like structures in uniaxial ferromagnets},
  author = {Bogdanov, A. and Hubert, A.},
  journal = {J. Magn. Magn. Mater.},
  volume = {195},
  number = {1},
  pages = {182--192},
  year = {1999},
  month = {apr},
  publisher = {Elsevier},
  doi = {10.1016/S0304-8853(98)01038-5},
  url = {https://www.sciencedirect.com/science/article/pii/S0304885398010385}
}

@article{Voinescu2020,
  title = {Hopf Solitons in Helical and Conical Backgrounds of Chiral Magnetic Solids},
  volume = {125},
  ISSN = {1079-7114},
  url = {http://dx.doi.org/10.1103/PhysRevLett.125.057201},
  DOI = {10.1103/physrevlett.125.057201},
  number = {5},
  journal = {Phys. Rev. Lett.},
  publisher = {American Physical Society (APS)},
  author = {Voinescu,  Robert and Tai,  Jung-Shen B. and Smalyukh,  Ivan I.},
  year = {2020},
  month = jul,
  pages     = {057201}
}

@article{Kuchkin2023,
  title = {Heliknoton in a film of cubic chiral magnet},
  volume = {11},
  ISSN = {2296-424X},
  url = {http://dx.doi.org/10.3389/fphy.2023.1201018},
  DOI = {10.3389/fphy.2023.1201018},
  journal = {Front. Phys.},
  publisher = {Frontiers Media SA},
  author = {Kuchkin,  Vladyslav M. and Kiselev,  Nikolai S. and Rybakov,  Filipp N. and Lobanov,  Igor S. and Bl\"{u}gel,  Stefan and Uzdin,  Valery M.},
  year = {2023},
  month = May,
  pages     = {1201018}
}

@article{Chen2026_1,
  title = {Electric‐Current‐Assisted Nucleation of Zero‐Field Hopfion Rings},
  volume = {38},
  ISSN = {1521-4095},
  url = {http://dx.doi.org/10.1002/adma.202523417},
  DOI = {10.1002/adma.202523417},
  number = {21},
  journal = {Adv. Mater.},
  publisher = {Wiley},
  author = {Chen,  Xiaowen and Song,  Dongsheng and Rybakov,  Filipp N. and Kiselev,  Nikolai S. and Li,  Long and Shi,  Wen and Wu,  Rui and Fu,  Xuewen and Eriksson,  Olle and Bl\"{u}gel,  Stefan and Dunin‐Borkowski,  Rafal E. and Du,  Haifeng and Zheng,  Fengshan},
  year = {2026},
  month = Mar,
  pages     = {2523417}
}

@article{Foster2019,
  title = {Two-dimensional skyrmion bags in liquid crystals and ferromagnets},
  volume = {15},
  ISSN = {1745-2481},
  url = {http://dx.doi.org/10.1038/s41567-019-0476-x},
  DOI = {10.1038/s41567-019-0476-x},
  number = {7},
  journal = {Nat. Phys.},
  publisher = {Springer Science and Business Media LLC},
  author = {Foster,  David and Kind,  Charles and Ackerman,  Paul J. and Tai,  Jung-Shen B. and Dennis,  Mark R. and Smalyukh,  Ivan I.},
  year = {2019},
  month = Apr,
  pages = {655–659}
}

@article{Kuchkin2025_v3,
  title = {Symmetry-governed dynamics of magnetic skyrmions under field pulses},
  volume = {8},
  ISSN = {2399-3650},
  url = {http://dx.doi.org/10.1038/s42005-024-01913-1},
  DOI = {10.1038/s42005-024-01913-1},
  number = {1},
  journal = {Commun. Phys.},
  publisher = {Springer Science and Business Media LLC},
  author = {Kuchkin,  Vladyslav M. and Barton-Singer,  Bruno and Bessarab,  Pavel F. and Kiselev,  Nikolai S.},
  year = {2025},
  month = Jan,
  pages     = {19}
}

@article{Chen2026_2,
  title = {Laser-induced nucleation of magnetic hopfions},
  volume = {22},
  ISSN = {1745-2481},
  url = {http://dx.doi.org/10.1038/s41567-026-03236-0},
  DOI = {10.1038/s41567-026-03236-0},
  number = {5},
  journal = {Nat. Phys.},
  publisher = {Springer Science and Business Media LLC},
  author = {Chen,  Xiaowen and Yang,  Donghai and Li,  Zefang and Guo,  Jiangteng and Wang,  Haixue and Hu,  Yue and Kuchkin,  Vladyslav M. and Savchenko,  Andrii S. and Zhang,  Huai and Ding,  Bei and Hou,  Zhipeng and Shi,  Wen and Rybakov,  Filipp N. and Eriksson,  Olle and Bl\"{u}gel,  Stefan and Han,  Yu and Dunin-Borkowski,  Rafal E. and Kiselev,  Nikolai S. and Fu,  Xuewen and Zheng,  Fengshan},
  year = {2026},
  month = Apr,
  pages = {736–744}
}

@article{Rybakov2019,
  title = {Chiral magnetic skyrmions with arbitrary topological charge},
  volume = {99},
  ISSN = {2469-9969},
  url = {http://dx.doi.org/10.1103/PhysRevB.99.064437},
  DOI = {10.1103/physrevb.99.064437},
  number = {6},
  journal = {Phys. Rev. B},
  publisher = {American Physical Society (APS)},
  author = {Rybakov,  Filipp N. and Kiselev,  Nikolai S.},
  year = {2019},
  month = Feb,
  pages     = {064437}
}

@article{Doring1968,
  author    = {W. D\"oring},
  title     = {Point Singularities in Micromagnetism},
  journal   = {J. Appl. Phys.},
  volume    = {39},
  pages     = {1006--1007},
  year      = {1968},
  doi       = {10.1063/1.1656143},
  url       = {http://link.aip.org/link/?JAP/39/1006/1}
}

@article{Liu2025,
  title = {Construction of hopfion crystals},
  author = {Hou, Wen-Tao and Xiang, Zhuoxian and Liu, Yizhou and Zang, Jiadong},
  journal = {Phys. Rev. Res.},
  volume = {7},
  issue = {4},
  pages = {L042036},
  numpages = {6},
  year = {2025},
  month = {Nov},
  publisher = {American Physical Society},
  doi = {10.1103/rr4k-x9m5},
  url = {https://link.aps.org/doi/10.1103/rr4k-x9m5}
}

@article{Kuchkin2022,
  title = {Homotopy transitions and 3D magnetic solitons},
  volume = {10},
  ISSN = {2166-532X},
  url = {http://dx.doi.org/10.1063/5.0097559},
  DOI = {10.1063/5.0097559},
  number = {7},
  journal = {APL Mater.},
  publisher = {AIP Publishing},
  author = {Kuchkin,  V. M. and Kiselev,  N. S.},
  year = {2022},
  month = {July},
  pages     = {071102}
}

@article{Zheng2021,
  title = {Magnetic skyrmion braids},
  volume = {12},
  ISSN = {2041-1723},
  url = {http://dx.doi.org/10.1038/s41467-021-25389-7},
  DOI = {10.1038/s41467-021-25389-7},
  number = {1},
  journal = {Nat. Commun.},
  publisher = {Springer Science and Business Media LLC},
  author = {Zheng,  Fengshan and Rybakov,  Filipp N. and Kiselev,  Nikolai S. and Song,  Dongsheng and Kovács,  András and Du,  Haifeng and Bl\"{u}gel,  Stefan and Dunin-Borkowski,  Rafal E.},
  year = {2021},
  month = {Sept},
  pages     = {5316}
}

@misc{mumax3_scripts,
  author       = {Vladyslav M. Kuchkin},
  title        = {MuMax3 simulation scripts},
  year         = {2026},
  version      = {1.0},
  publisher    = {Zenodo},
  url          = {https://10.5281/zenodo.24245757}
}

@article{Kuchkin2022_v2,
  title = {Thermal generation of droplet soliton in chiral magnet},
  volume = {105},
  ISSN = {2469-9969},
  url = {http://dx.doi.org/10.1103/PhysRevB.105.184403},
  number = {18},
  journal = {Phys. Rev. B},
  publisher = {American Physical Society (APS)},
  author = {Kuchkin,  Vladyslav M. and Bessarab,  Pavel F. and Kiselev,  Nikolai S.},
  year = {2022},
  month = May,
  pages     = {184403}
}

@article{Saji2025,
  title = {Statics and dynamics of twisted skyrmion tubes in frustrated magnets},
  volume = {112},
  ISSN = {2469-9969},
  url = {http://dx.doi.org/10.1103/tz4v-s5m5},
  DOI = {10.1103/tz4v-s5m5},
  number = {13},
  journal = {Phys. Rev. B},
  publisher = {American Physical Society (APS)},
  author = {Saji,  Carlos and Saavedra,  Eduardo and Carvalho-Santos,  Vagson L. and Nunez,  Alvaro S. and Troncoso,  Roberto E.},
  year = {2025},
  month = Oct,
  pages     = {134414}
}

@article{Leonov2016,
  title = {Chiral Surface Twists and Skyrmion Stability in Nanolayers of Cubic Helimagnets},
  volume = {117},
  ISSN = {1079-7114},
  url = {http://dx.doi.org/10.1103/PhysRevLett.117.087202},
  DOI = {10.1103/physrevlett.117.087202},
  number = {8},
  journal = {Phys. Rev. Lett.},
  publisher = {American Physical Society (APS)},
  author = {Leonov,  A. O. and Togawa,  Y. and Monchesky,  T. L. and Bogdanov,  A. N. and Kishine,  J. and Kousaka,  Y. and Miyagawa,  M. and Koyama,  T. and Akimitsu,  J. and Koyama,  Ts. and Harada,  K. and Mori,  S. and McGrouther,  D. and Lamb,  R. and Krajnak,  M. and McVitie,  S. and Stamps,  R. L. and Inoue,  K.},
  year = {2016},
  month = Aug,
  pages     = {087202}
}

@article{Tang2021,
  title = {Magnetic skyrmion bundles and their current-driven dynamics},
  volume = {16},
  ISSN = {1748-3395},
  url = {http://dx.doi.org/10.1038/s41565-021-00954-9},
  DOI = {10.1038/s41565-021-00954-9},
  number = {10},
  journal = {Nat. Nanotechnol.},
  publisher = {Springer Science and Business Media LLC},
  author = {Tang,  Jin and Wu,  Yaodong and Wang,  Weiwei and Kong,  Lingyao and Lv,  Boyao and Wei,  Wensen and Zang,  Jiadong and Tian,  Mingliang and Du,  Haifeng},
  year = {2021},
  month = Aug,
  pages = {1086–1091}
}

@article{Zhang2024,
  title = {Stable skyrmion bundles at room temperature and zero magnetic field in a chiral magnet},
  volume = {15},
  ISSN = {2041-1723},
  url = {http://dx.doi.org/10.1038/s41467-024-47730-6},
  DOI = {10.1038/s41467-024-47730-6},
  number = {1},
  journal = {Nat. Commun.},
  publisher = {Springer Science and Business Media LLC},
  author = {Zhang,  Yongsen and Tang,  Jin and Wu,  Yaodong and Shi,  Meng and Xu,  Xitong and Wang,  Shouguo and Tian,  Mingliang and Du,  Haifeng},
  year = {2024},
  month = Apr,
  pages     = {3391}
}

@article{Tai2019,
  title = {Three-dimensional crystals of adaptive knots},
  volume = {365},
  ISSN = {1095-9203},
  url = {http://dx.doi.org/10.1126/science.aay1638},
  DOI = {10.1126/science.aay1638},
  number = {6460},
  journal = {Science},
  publisher = {American Association for the Advancement of Science (AAAS)},
  author = {Tai,  Jung-Shen B. and Smalyukh,  Ivan I.},
  year = {2019},
  month = {Sept},
  pages = {1449–1453}
}

@article{Rybakov2022,
  title = {Magnetic hopfions in solids},
  volume = {10},
  ISSN = {2166-532X},
  url = {http://dx.doi.org/10.1063/5.0099942},
  DOI = {10.1063/5.0099942},
  number = {11},
  journal = {APL Mater.},
  publisher = {AIP Publishing},
  author = {Rybakov,  Filipp N. and Kiselev,  Nikolai S. and Borisov,  Aleksandr B. and D\"{o}ring,  Lukas and Melcher,  Christof and Bl\"{u}gel,  Stefan},
  year = {2022},
  month = Nov,
  pages     = {111113}
}

\newpage
\clearpage
\onecolumngrid

\newpage
\begin{center}
   \textbf{Supplemental Material for ``Higher‑order magnetic hopfions in uniaxial chiral magnets''}\newline 
\end{center}

\renewcommand{\theequation}{S\arabic{equation}}
\setcounter{equation}{0}   

\renewcommand{\thefigure}{S\arabic{figure}}
\setcounter{figure}{0}

\section{Calculation of the Hopf index}
\label{sec:hopf-charge}

We calculate the Hopf index using the method
introduced in Ref.~\cite{Chen2026_1}.
This method allows us to determine the topological
invariant of localized magnetic textures
with nonuniform boundary magnetization.
As shown in Ref.~\cite{Chen2026_1},
the third relative homotopy group satisfies
\begin{equation}
\pi_3\left(
\mathbb{S}^2,
\mathbb{S}^2\setminus\bigcup_{i=1}^{n}X_i
\right)
\simeq\mathbb{Z}, \qquad n\geq 1,
\label{relative_homotopy}
\end{equation}
where $X_i$ are disjoint regions excluded
from the sphere of magnetization directions.
For a hopfion embedded in the conical phase,
the boundary magnetization follows a circle
on this sphere.
Allowing for deviations from the ideal conical
background, the corresponding boundary subspace
can be represented by a sphere with two excluded
regions.
Therefore, the topological classification
remains characterized by an integer Hopf index $H$.

\begin{figure*}[htbp]
\centering
\includegraphics[width=18cm]{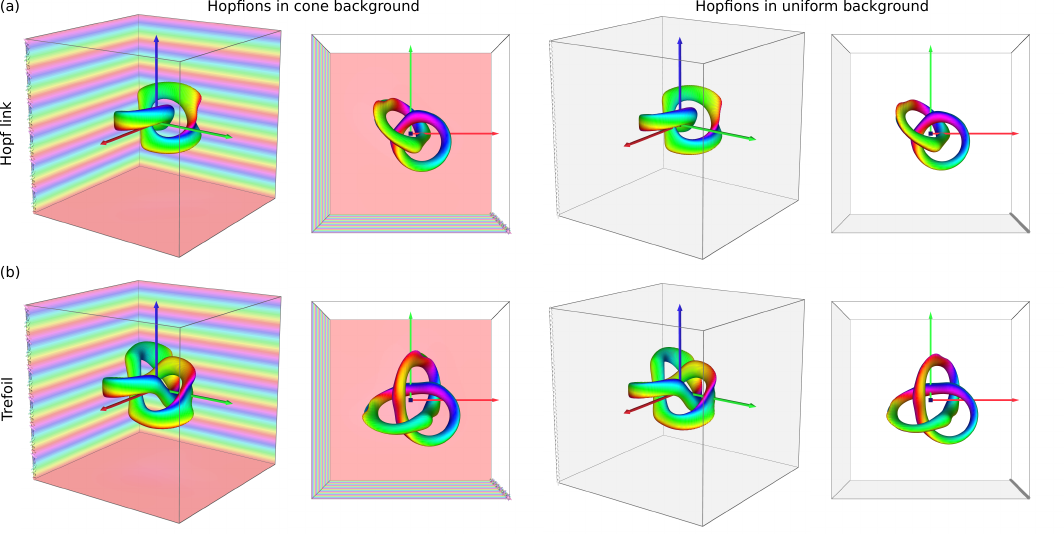}
\caption{\textbf{Effect of the auxiliary projection on magnetic hopfions.}
(a) Hopf link with Hopf index $H=0$.
(b) Trefoil knot with Hopf index $H=1$.
For each configuration, the left panel shows
the hopfion embedded in the conical background
after energy minimization.
The right panel shows the same configuration
after applying the auxiliary projection
defined in Eq.~\eqref{hopf_projection}.
}
\label{fig:hopf_projection}
\end{figure*}

The calculation involves two steps.
First, we apply an auxiliary projection
$\mathbf{m}(\mathbf{r})\mapsto
\mathbf{n}(\mathbf{r})$,
which preserves the Hopf index while transforming
the boundary magnetization into a uniform state.
Following Ref.~\cite{Chen2026_1},
we use the transformation
\begin{equation}
\mathbf{n}=
\begin{pmatrix}
2\gamma m_x\\
2\gamma m_y\\
1-2\gamma^2(1-m_z)(1+\mu)
\end{pmatrix},
\label{hopf_projection}
\end{equation}
where
\begin{equation}
\gamma=
\begin{cases}
\displaystyle
\sqrt{\frac{\mu-m_z}{(1-m_z)(1+\mu)^2}},
& m_z<\mu,\\[6pt]
0, & m_z\geq\mu.
\end{cases}
\label{hopf_gamma}
\end{equation}
The parameter $\mu=\mu(\mathbf{r})$ varies
from $+1$ inside the calculation domain
to $-1$ at its boundary.
In normalized coordinates $r_i\in[-1,1]$,
the projection function is defined as
\begin{equation}
\mu(\mathbf r)=
1-2\min\left(1,r_x^2+r_y^2+r_z^2\right).
\label{hopf_mu}
\end{equation}
This transformation maps the boundary
magnetization onto the north pole
$\mathbf{n}=(0,0,1)$ without changing
the topological invariant.
It can also be viewed as mapping the
nonuniform magnetic background onto a
uniform reference state at the boundary.
A similar approach was used in
Ref.~\cite{Zheng2023} to calculate
the topological indices of Hopfion rings
by aligning the magnetization at the
lateral boundaries of the calculation domain,
while retaining periodic boundary conditions
along $z$.
In the present work, the projection is applied
in all three spatial directions, yielding
uniform magnetization on the entire boundary
of the calculation domain.
Its effect on the Hopf link and trefoil
configurations is illustrated in
Figs.~\ref{fig:hopf_projection}(a) and
\ref{fig:hopf_projection}(b).

In the second step, we calculate the Hopf
index of the transformed field using
the Whitehead integral:
\begin{equation}
H=-\frac{1}{16\pi^2}
\int_{\Omega}\mathrm{d}^3r\,
\mathbf{F}\cdot\mathbf{A},
\label{Hopf}
\end{equation}
where the curvature vector is defined by
\begin{equation}
F_i=\frac12\epsilon_{ijk}
\mathbf{n}\cdot
\left(
\partial_j\mathbf{n}
\times
\partial_k\mathbf{n}
\right),
\label{hopf_curvature}
\end{equation}
and the vector potential satisfies
$\nabla\times\mathbf{A}=\mathbf{F}$.
We calculate $\mathbf{F}$ using the
Berg--L\"uscher method~\cite{BERG1981412}
and obtain the vector potential by numerical
integration~\cite{Rybakov2022}:
\begin{equation}
    \mathbf{A} = \left(-\int_{0}^{y}F_{z}\mathrm{d}y, 0, \int_{0}^{y}F_{x}\mathrm{d}y\right).
    \label{A}
\end{equation}
Here, $\Omega=[0,L]^3$ is the calculation domain,
and the projected magnetization field
$\mathbf{n}$ is uniform on its boundary,
$\mathbf{n}|_{\partial\Omega}=(0,0,1)$.
Alternative numerical methods for calculating
the Hopf index are discussed in
Ref.~\cite{Knapman2025}.
An alternative approach to classifying magnetic
textures with nonuniform boundary conditions
has recently been proposed, in which the
Whitehead integral is evaluated directly,
without an auxiliary projection \eqref{hopf_projection}, yielding
noninteger values~\cite{Azhar2026}.

Finally, it is worth noting that changing the system chirality, $\mathbf{m}\cdot\left(\nabla \times \mathbf{m}\right) \mapsto - \mathbf{m}\cdot\left(\nabla \times \mathbf{m}\right)$ is mathematically equivalent to the space inversion operation, $\mathbf{r}\mapsto -\mathbf{r}$.
As can be easily seen, the inversion does not change the gyro-vector because it contains the derivative with respect to spatial variables twice, and the minus signs cancel out. 
However, from \eqref{A}, we see that inversion changes $\mathbf{A}\mapsto -\mathbf{A}$, and thus the Hopf index flips sign as well, $H \mapsto -H$, in the system of opposite chirality.

\section{Finite-difference coefficients}
\label{sec:fin-diff}

\begin{figure}[htbp]
\centering
\includegraphics[width=6.7cm]{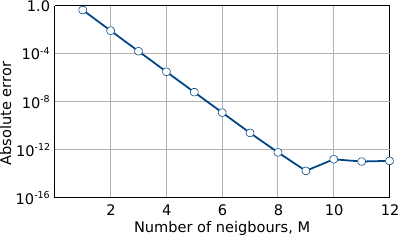}
\caption{Error in the calculation of the cone phase energy for $\Delta = 3/64$.}
\label{FigS1}
\end{figure}

In numerical minimization of the chiral magnet Hamiltonian, we effectively discretize exchange and DMI energy density terms as:
\begin{eqnarray}
    &&w_\mathrm{ex}=\dfrac{2}{\Delta^{2}}\sum_{s=1}^{M}J_{s}\sum_{ijk}\left[3-\mathbf{m}_{ijk}\cdot\left(\mathbf{m}_{i-s,jk}+\mathbf{m}_{i,j-s,k}+\mathbf{m}_{ij,k-s}\right)\right],\label{ex_num}\\
    &&w_\mathrm{dmi} = \dfrac{1}{\Delta}\sum_{s=1}^{M}D_{s}\!\!\sum_{ijk} \mathbf{m}_{ijk}\!\cdot\!\left[\mathbf{m}_{i-s,jk}\times\mathbf{e}_{x} + \mathbf{m}_{i,j-s,k}\times\mathbf{e}_{y} + \mathbf{m}_{ij,k-s}\times\mathbf{e}_{z}\right],
\end{eqnarray}
where $M$ is the number of neighbors used, which provides a scheme of accuracy $(2M)$; $\mathbf{e}_{x}=(1,0,0)$, $\mathbf{e}_{y}=(0,1,0)$ and $\mathbf{e}_{z}=(0,0,1)$ are basis vectors of $\mathbb{R}^{3}$; $\Delta $ is the inverse of the number of nodes per spiral period $L_{D}$, which we set to $3/64 = 0.046875$, corresponding to about $21$ spins per $L_{D}$.
To reliably simulate large systems with $256^{3}$ and $512^{3}$ spins for a fixed $\Delta$, we use higher-order finite-difference schemes with order $M$.
By comparing the cone-phase energy with the theoretical value, we estimate the absolute error for different values of $M$ [see Fig.~\ref{FigS1}]. 
To reach machine precision (double in C++), choosing $M=9$ is sufficient. Utilizing higher values of $M$ does not improve accuracy. 
For $M=4$, the error is about $10^{-6}$, which coincides with the single-precision error. 
As for some calculations, we rely on Mumax3 software, which uses single-precision floating-point numbers; we use the scheme with $M=4$ for all simulations presented in the paper.
We compute the exact values of the coefficients $J_{s}$ and $D_{s}$ according to Ref.~\cite{Fornberg1988} and provide them in Table~\ref{tab:Js_Ds}.

\begin{table}[htbp]
    \centering
\renewcommand{\arraystretch}{2.0}          
\setlength{\tabcolsep}{10pt}               
\begin{tabular}{|*{5}{>{\centering\arraybackslash}c|}}
\hline
\rule[-0.4cm]{0pt}{1.1cm}$s$ & $1$ & $2$ & $3$ & $4$ \\
\hline
\rule[-0.4cm]{0pt}{1.2cm}$J_s$ &
$\dfrac{8}{5}$ &
$-\dfrac{1}{5}$ &
$\dfrac{8}{315}$ &
$-\dfrac{1}{560}$ \\
\hline
\rule[-0.4cm]{0pt}{1.2cm}$D_s$ &
$\dfrac{8}{5}$ &
$-\dfrac{2}{5}$ &
$\dfrac{8}{105}$ &
$-\dfrac{1}{140}$ \\
\hline
\end{tabular}
\captionof{table}{Values of $J_s$ and $D_s$ for different values of $s$ for $M=4$.}
\label{tab:Js_Ds}
\end{table}

\section{Interpolation in cylindrical coordinates}
\label{sec:Interpolation}

To obtain the magnetization field on a Cartesian grid,
we interpolate each component of $\mathbf{m}$
separately using the cylindrical-coordinate representation.
The resulting magnetization vectors are then normalized
to satisfy $|\mathbf{m}|=1$.
For any magnetization component
$m(r,\phi,z)$ defined at the grid nodes
$(i\delta r,j\delta\phi,k\delta z)$,
we use the following interpolation function:
\begin{equation}
F\left(r, \phi, z\right)=f_{0}\left(r, z\right) + f_{1}\left(r, z\right)\cos\phi + f_{2}\left(r, z\right)\sin\phi,\quad f_{l}=a_{l} + b_{l}r + c_{l}z + d_{l}rz, \quad l=\overline{0,2}\label{interp_cylinder},
\end{equation}
where $r\in[i\delta r, \left(i+1\right)\delta r]$,  $\phi\in[j\delta \phi, \left(j+1\right)\delta \phi]$,  $z\in[k\delta z, \left(k+1\right)\delta z]$, and 12 coefficients $a_{l}, b_{l}, c_{l}, d_{l}$ are to be determined. 
Denoting $r=i \delta r$, $\phi = j \delta \phi$ and $z = k \delta z$, this can be done by solving the following system of equations:
\begin{eqnarray}
F\left(r,\phi,z\right)=m_{0},&F\left(r+\delta r,\phi,z\right)=m_{1},&F\left(r,\phi+\delta\phi,z\right)=m_{2},\\F\left(r+\delta r,\phi+\delta\phi,z\right)=m_{3},&F\left(r,\phi-\delta\phi,z\right)=m_{4},&F\left(r+\delta r,\phi-\delta\phi,z\right)=m_{5},\\F\left(r,\phi,z+\delta z\right)=m_{6},&F\left(r+\delta r,\phi,z+\delta z\right)=m_{7},&F\left(r,\phi+\delta\phi,z+\delta z\right)=m_{8},\\F\left(r+\delta r,\phi+\delta\phi,z+\delta z\right)=m_{9},&F\left(r,\phi-\delta\phi,z+\delta z\right)=m_{10},&F\left(r+\delta r,\phi-\delta\phi,z+\delta z\right)=m_{11}.  
\end{eqnarray}
This system can be easily solved as follows. First introducing the matrix,
\begin{equation}
    A=\left(\begin{array}{ccc}
1 & \cos\phi & \sin\phi\\
1 & \cos\left(\phi+\delta\phi\right) & \sin\left(\phi+\delta\phi\right)\\
1 & \cos\left(\phi-\delta\phi\right) & \sin\left(\phi-\delta\phi\right)
\end{array}\right),
\end{equation}
we can obtain the following:
\begin{equation}
    \left(\begin{array}{c}
b_{0}+d_{0}z\\
b_{1}+d_{1}z\\
b_{2}+d_{2}z
\end{array}\right)=\dfrac{1}{\delta r}A^{-1}\left(\begin{array}{c}
m_{1}-m_{0}\\
m_{3}-m_{2}\\
m_{5}-m_{4}
\end{array}\right),\quad\left(\begin{array}{c}
b_{0}+d_{0}\left(z+\delta z\right)\\
b_{1}+d_{1}\left(z+\delta z\right)\\
b_{2}+d_{2}\left(z+\delta z\right)
\end{array}\right)=\dfrac{1}{\delta r}A^{-1}\left(\begin{array}{c}
m_{7}-m_{6}\\
m_{9}-m_{8}\\
m_{11}-m_{10}
\end{array}\right),
\end{equation}
from which coefficients $b_{l}, d_{l}$ are readily found.
Secondly, introducing the matrix
\begin{equation}
    B=\left(\begin{array}{cc}
\cos\left(\phi+\delta\phi\right)-\cos\phi & \sin\left(\phi+\delta\phi\right)-\sin\phi\\
\cos\left(\phi-\delta\phi\right)-\cos\phi & \sin\left(\phi-\delta\phi\right)-\sin\phi
\end{array}\right),
\end{equation}
we can get 
\begin{eqnarray}
    &\left(\begin{array}{c}
a_{1}+c_{1}z\\
a_{2}+c_{2}z
\end{array}\right)=B^{-1}\left(\begin{array}{c}
m_{3}-m_{1}\\
m_{5}-m_{1}
\end{array}\right)-\left(r+\delta r\right)\left(\begin{array}{c}
b_{1}+d_{1}z\\
b_{2}+d_{2}z
\end{array}\right),\nonumber \\
&\left(\begin{array}{c}
a_{1}+c_{1}\left(z+\delta z\right)\\
a_{2}+c_{2}\left(z+\delta z\right)
\end{array}\right)=B^{-1}\left(\begin{array}{c}
m_{9}-m_{7}\\
m_{11}-m_{7}
\end{array}\right)-\left(r+\delta r\right)\left(\begin{array}{c}
b_{1}+d_{1}\left(z+\delta z\right)\\
b_{2}+d_{2}\left(z+\delta z\right)
\end{array}\right),
\end{eqnarray}
which allows us to determine $a_{1}, a_{2}, c_{1}$ and $c_{2}$.
Finally, we get $c_{0}$ and $a_{0}$ as follows,
\begin{eqnarray}
    &c_{0}=\dfrac{m_{6}-m_{0}}{\delta z}-d_{0}r-\left(c_{1}+d_{1}r\right)\cos\phi-\left(c_{2}+d_{2}r\right)\sin\phi,\nonumber\\
    &a_{0}=m_{0}-b_{0}r-c_{0}z-d_{0}rz-f_{1}\left(r,z\right)\cos\phi-f_{2}\left(r,z\right)\sin\phi.
\end{eqnarray}

The braiding functions used for constructing initial states are provided in Table~\ref{tab:braiding_functions} and \textit{Mathematica} scripts are available in Ref.~\cite{mumax3_scripts}.

\begin{table}[htbp]
    \centering
\renewcommand{\arraystretch}{2.0}          
\setlength{\tabcolsep}{10pt}               
\begin{tabular}{|*{5}{>{\centering\arraybackslash}c|}}
\hline
\rule[-0.4cm]{0pt}{1.1cm}Braiding word & $(f_{i}, g_{i})$  \\
\hline
\rule[-0.4cm]{0pt}{1.2cm}$\sigma_{1}^{n}$ &
$\begin{array}{cc}
     \left(f_{1}, g_{1}\right)=R\left(\cos \dfrac{\pi nz}{L}, \sin \dfrac{\pi nz}{L}\right),  \quad
     \left(f_{2}, g_{2}\right)=-R\left(\cos \dfrac{\pi nz}{L}, \sin \dfrac{\pi nz}{L}\right). 
\end{array}$  \\
\hline
\rule[-0.4cm]{0pt}{1.2cm}$\left(\sigma_{2}\sigma_{1}\right)^{n}$ &
$\begin{array}{cc}
     \left(f_{1}, g_{1}\right)=\dfrac{2R}{\sqrt{3}}\left(\cos \dfrac{2\pi nz}{3L}, \sin \dfrac{2\pi nz}{3L} \right),  \\
     \left(f_{2}, g_{2}\right)=\dfrac{2R}{\sqrt{3}}\left(\cos \left[\dfrac{2\pi nz}{3L} + \dfrac{2\pi}{3}\right], \sin \left[\dfrac{2\pi nz}{3L} + \dfrac{2\pi}{3}\right] \right), \\ 
     \left(f_{3}, g_{3}\right)=\dfrac{2R}{\sqrt{3}}\left(\cos \left[\dfrac{2\pi nz}{3L} + \dfrac{4\pi}{3}\right], \sin \left[\dfrac{2\pi nz}{3L} + \dfrac{4\pi}{3}\right] \right).
\end{array}$  \\
\hline
\rule[-0.4cm]{0pt}{1.2cm}$\left(\sigma_{1}\sigma_{2}\sigma_{1}\right)^{n}$ &
$\begin{array}{cc}
     \left(f_{1}, g_{1}\right)=2R\left(\cos \dfrac{\pi nz}{L}, \sin \dfrac{\pi nz}{L}\right),  \quad
     \left(f_{2}, g_{2}\right)=-2R\left(\cos \dfrac{\pi nz}{L}, \sin \dfrac{\pi nz}{L}\right),\\
     \left(f_{3}, g_{3}\right) = \left(0,0\right).
\end{array}$  \\
\hline
\rule[-0.4cm]{0pt}{1.2cm}$\left(\sigma_{2}\sigma_{1}\sigma_{3}\sigma_{2}\right)^{n}\left(\sigma_{3}\sigma_{1}\right)^{m}$ &
$\begin{array}{cc}
     \left(f_{1},g_{1}\right)=-R\left(\cos\dfrac{\pi mz}{L}-7+2\cos\dfrac{\pi nz}{L},\sin\dfrac{\pi mz}{L}+2\sin\dfrac{\pi nz}{L}\right),  \\
     \left(f_{2},g_{2}\right)=R\left(\cos\dfrac{\pi mz}{L}+7-2\cos\dfrac{\pi nz}{L},\sin\dfrac{\pi mz}{L}-2\sin\dfrac{\pi nz}{L}\right),\\
     \left(f_{3},g_{3}\right)=R\left(-\cos\dfrac{\pi mz}{L}+7+2\cos\dfrac{\pi nz}{L},-\sin\dfrac{\pi mz}{L}+2\sin\dfrac{\pi nz}{L}\right),\\
     \left(f_{4},g_{4}\right)=R\left(\cos\dfrac{\pi mz}{L}+7+2\cos\dfrac{\pi nz}{L},\sin\dfrac{\pi mz}{L}+2\sin\dfrac{\pi nz}{L}\right).
\end{array}$  \\
\hline
\end{tabular}
\captionof{table}{Braiding functions}
\label{tab:braiding_functions}
\end{table}

\section{Extended set of various hopfions}
\label{sec:hopfion_diversity}

\begin{figure*}[htbp]
\centering
\includegraphics[width=18cm]{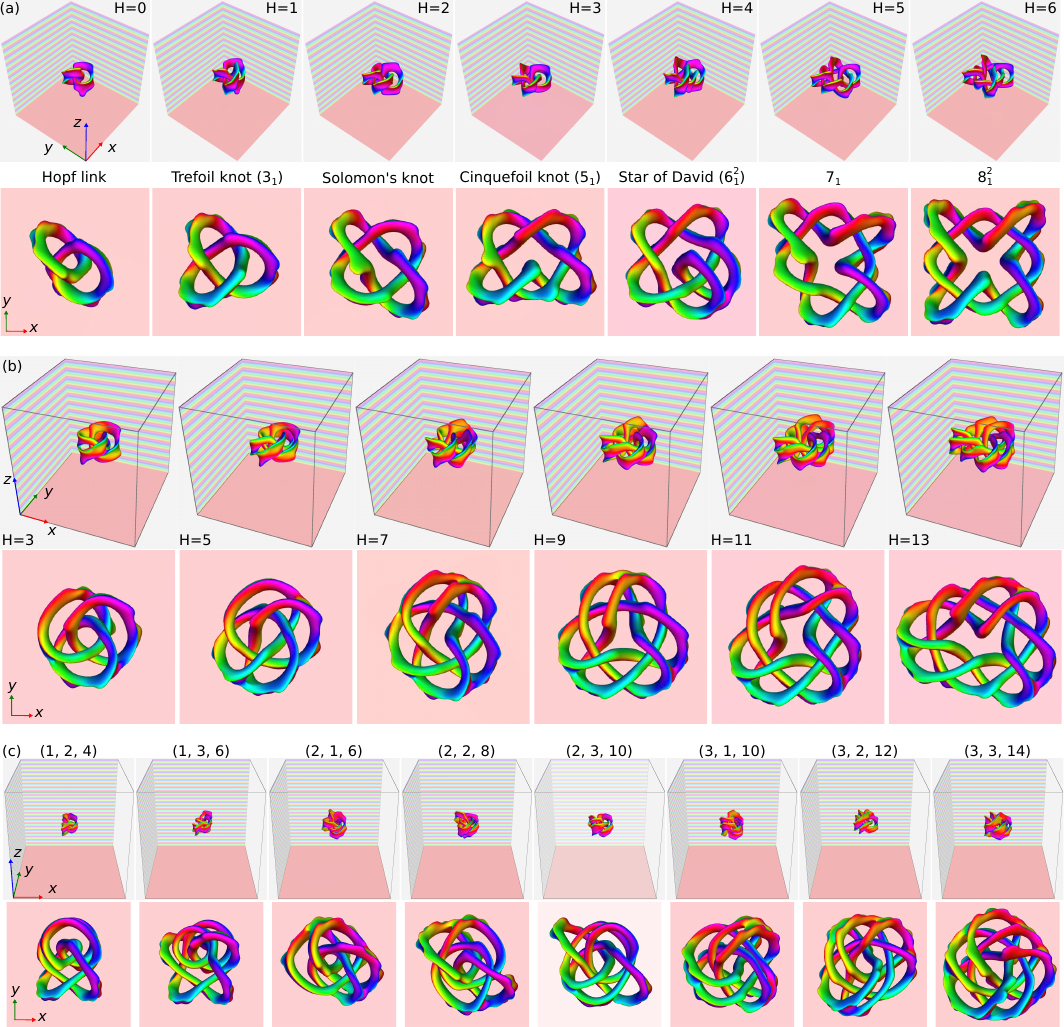}
\caption{\textbf{Extended set of hopfions}. (a), (b) and (c) contain stable hopfions obtained with braiding words $\left(\sigma_{1}\right)^{n}$, $n=\overline{2,8}$, $\left(\sigma_{2}\sigma_{1}\right)^{n}$, $n=\overline{3,8}$, and  $\left(\sigma_{2}\sigma_{1}\sigma_{3}\sigma_{2}\right)^{n}\left(\sigma_{3}\sigma_{1}\right)^{m}$, respectively. 
Hopf charges are provided in (a) and (b); the values of $(n, m, H)$ are given in (c). 
Upper raw corresponds to the hopfions in full system, while lower rows contain zoom-in top view.}
\label{FigS2}
\end{figure*}

In Fig.~\ref{FigS2}(a)-(c), we provide more examples of stable hopfions at $(u,h)=(0.2, 0.43)$, the only case in (c) with $(n, m, H)=(2, 3, 10)$ is given at $(u,h)=(0.23, 0.43)$.
The 3D view shows the full system used in simulations $(12l)^{3}$ split into $256^3$ cuboids for (a) and (b), and $(24l)^{3}$ split into $512^3$ cuboids for case (c).
Energies of these states are given in Fig. 3(e) in the main text.

\end{document}